\documentclass[sigconf]{acmart}

\AtBeginDocument{%
  \providecommand\BibTeX{{%
    \normalfont B\kern-0.5em{\scshape i\kern-0.25em b}\kern-0.8em\TeX}}}

\usepackage{subcaption}
\usepackage{enumitem}

\copyrightyear{2021}
\acmYear{2021}
\acmBooktitle{CHI Conference on Human Factors in Computing Systems Extended Abstracts (CHI '21 Extended Abstracts), May 8--13, 2021, Yokohama, Japan}
\acmDOI{10.1145/3411763.3451643}
\acmISBN{978-1-4503-8095-9/21/05}

\begin{document}

\title{Facilitating Asynchronous Participatory Design of Open Source Software: Bringing End Users into the Loop}

\author{Jazlyn Hellman}
\email{jazlyn.hellman@mail.mcgill.ca}
\affiliation{%
 \institution{McGill University}
 \city{Montreal}
 \state{Quebec}
 \country{Canada}}

\author{Jinghui Cheng}
\email{jinghui.cheng@polymtl.ca}
\affiliation{%
 \institution{Polytechnique Montreal}
 \city{Montreal}
 \state{Quebec}
 \country{Canada}}

\author{Jin L.C. Guo}
\email{jguo@cs.mcgill.ca}
\affiliation{%
 \institution{McGill University}
 \city{Montreal}
 \state{Quebec}
 \country{Canada}}

\begin{abstract}
    As open source software (OSS) becomes increasingly mature and popular, there are significant challenges with properly accounting for usability concerns for the diverse end users. Participatory design, where multiple stakeholders collaborate on iterating the design, can be an efficient way to address the usability concerns for OSS projects. However, barriers such as a code-centric mindset and insufficient tool support often prevent OSS teams from effectively including end users in participatory design methods. This paper proposes preliminary contributions to this problem through the user-centered exploration of (1) a set of design guidelines that capture the needs of OSS participatory design tools, (2) two personas that represent the characteristics of OSS designers and end users, and (3) a low-fidelity prototype tool for end user involvement in OSS projects. This work paves the road for future studies about tool design that would eventually help improve OSS usability.
\end{abstract}



\maketitle

\section{Introduction}
With the increasing maturity and popularity of open source software (OSS), ongoing efforts to increase the usability of the software developed under the open source model are growing in importance~\cite{cheryl_study}.
Usability refers to the attributes which determine the ease, error-prevention, efficiency, and pleasantness for an end user when interacting with a software ~\cite{cheng_guo_uxissues, nielsen_book}.
In traditional software development, usability concerns rest in the hands of design teams performing activities that involve various stakeholders in the iterative lifespan of a project to achieve a system design that responds to the needs of the end users. Much of this practice relies on the principles of participatory design, which places emphasis on the involvement and collaboration of all stakeholders, especially end users~\cite{Muller1993}.
While user participation in the design process is vital for achieving successful usability of an OSS, it is often pushed aside as teams adapt to asynchronous, remote working and focus on a project's code and functionality~\cite{FoSS_usability, cheryl_study}.

Take for example GitHub, one of the most successful OSS hosting and development platform. Due to its affinity for supporting various asynchronous activities, GitHub has enabled a vibrant community for remote collaboration and development of OSS projects. While there are many tools that can potentially be used for monitoring usability interests on GitHub OSS's, such as Issue tracking, there is currently no mature method for end user community collaboration~\cite{cheryl_study, argulens, FoSS_usability}.
Consequently, if any design decisions are made through input from end users, there are difficulties with properly communicating these decisions to other team members due to a lack of design artifacts clearly documenting the process.
Many OSS systems, as a result, are designed without direct input from end users. While prior work explored the pitfalls of end user community involvement in OSS design~\cite{cheryl_study, Andreasen2006, Raza2010a, FoSS_usability}, as of yet, there have been little contribution to the open source community offering a solution to this problem.

In this paper, we address this gap by exploring the design of a low-fidelity prototype to facilitate asynchronous participatory design in large-scale OSS projects. Our design decisions were originally informed by related work on challenges of addressing OSS usability from the developers' perspective and later iterated through preliminary user studies with both OSS designers and end users. Through this study, we contribute an initial set of personas and design guidelines for designing platforms that promote end users' participation in asynchronous, participatory OSS design, as well as a preliminary tool to achieve this primary design objective. This work lays the foundation for future studies of more full-fledged tools that would eventually improve OSS usability.

\begin{table*}[t]
    \small
    \caption{Design Guidelines for OSS Designers (* indicates challenges gathered during user studies.)}
    \begin{tabular}{lll}
    \hline
    \textbf{No.} & \textbf{Description} & \textbf{Targeted Challenges} \\ \hline
    D1 & OSS-D are able to easily engage with end users and understand their needs. & User Needs~\cite{cheryl_study}  \\
    D2 & The system integrates design practices into an existing development pipeline. & Mindset~\cite{cheryl_study}; Pipeline*  \\
    D3 & OSS-D are able to generate artifacts from interacting with end users. & User Diversity~\cite{cheryl_study}; Tracing Artifacts*  \\
    D4 & Design for a project can happen at any stage of an OSS project's lifecycle. & Development~\cite{cheryl_study}  \\
    D5 & The system encourages community-wide involvement in an OSS project. & Mindset~\cite{cheryl_study}; User Needs~\cite{cheryl_study}  \\ \hline
    \end{tabular}
    \label{tab:designer_goals}
\end{table*}

\begin{table*}[t]
    \small
    \caption{Design Guidelines for OSS End Users (* indicates challenges gathered during user studies.)}
    \begin{tabular}{lll}
        \hline
        \textbf{No.} & \textbf{Description} & \textbf{Targeted Challenges}\\ \hline
        EU1 & OSS-EU should easily learn how to collaborate on a project. & Inclusivity*, Learnability* \\
        EU2 & OSS-EU should interact with the OSS team through asynchronous discussions. & Development~\cite{cheryl_study}, Mindset~\cite{cheryl_study}  \\
        EU3 & OSS-EU should be able to draw attention to current usability issues they face. & Mindset~\cite{cheryl_study}, User Needs~\cite{cheryl_study}  \\
        EU4 & OSS-EU contributions should be recognized. & Transparency*, Recognition* \\
        EU5 & OSS-EU should feel motivated to collaborate on a project. & Inclusivity*, Transparency* \\ \hline
    \end{tabular}
    \label{tab:enduser_goals}
\end{table*}

\section{Background and Related Work}
Usability in the context of OSS has been a topic of interest for decades and is constantly shifting in nature. OSS has certainly garnered certain reputation for being `by developers for developers'; prior work explored by Nichols and Twidale (2002) has stated a sense of `elitism' among the OSS developers where pride was drawn from creating hard-to-learn but powerful products~\cite{Nichols_Twidale_2003}.
As the OSS community has grown and diversified in functionality priorities, general practices, software adoption, and user base, much work has sought to understand these shifting dynamics and power relations between OSS developers, designers, and end users~\cite{Nichols_Twidale_2003, FoSS_usability, cheng_guo_uxissues, oss_power}.

Currently, many OSS usability practices and discussions take place in an ad-hoc manner and, especially on OSS platforms such as GitHub (www.github.com), within a repository's issue tracking system~\cite{FoSS_usability, cheng_guo_uxissues, argulens}.
Cheng and Guo~\cite{cheng_guo_uxissues} found usability specific issue threads on GitHub to be lengthy and contain over-generalized assumptions. Wang et al.~\cite{cheryl_study} expressed the need for a shift to a user-centric mindset amongst OSS practitioners along with a standardized way to include end users in the OSS project. However, there is little previous work that focused on tool design for alleviating these problems. We address this gap by exploring tools for end user involvement and collaboration.

Moreover, most previous work has focused on the practices and challenges of the OSS developers~\cite{Nichols_Twidale_2003, cheryl_study, FoSS_usability}; little investigation has been conducted from the perspective of `non-technical' end users. Falling into a similar pitfall is the recently released GitHub Discussions.
~\cite{ushev_2020, discussions}.
This feature is designed to replace the lengthy conversations currently taking place on issues and pull requests by creating a dedicated community conversation space within the GitHub repository.
However, this feature continues to place the focus on the developers of the OSS community and not others.
While features such as `GitHub Discussions' could potentially be used for usability discussions, the problem remains that there is still no dedicated place for handling end user involvement and participatory design efforts nor any concrete artifacts being created to document the consequent design decisions. Our work attempts to address these gaps through a user-centered design exploration of a OSS tool for participatory design.

\section{Design Process}
In this section, we detail how we explored the tool design, presenting the target user groups, the usability goals and heuristics, and the ideation and overall design process.

\subsection{Target User Groups}
To understand the context of when and where issues will be faced during asynchronous participatory design as well as the nature of those issues, we defined two target user groups of our intended system: (1) OSS contributors focusing on the design aspects of OSS projects (\textbf{OSS-D}) and (2) non-technical end users of OSS projects (\textbf{OSS-EU}).

We selected OSS contributors focused on design (referred to as OSS designers from here on) as the first User Group because:
(1) these individuals are the ones most impacted by the lack of standardization in OSS design practices;
(2) they would have the most insight on current barriers facing participatory design integration in asynchronous OSS work.
In regards to the second user group, we are interested in the case where the end users of the OSS are not the technical contributors to the OSS project, but rather the direct user of the end product.
These target audiences were used to (1) direct our design decisions and (2) guide participant selection for the user studies (see Section \ref{section:user_study}).

\subsection{Design Guidelines}
\label{section:designguidelines}
Once the target audiences were finalized, we created a set of design guidelines to support the exploration of the system design. Defining these guidelines was an iterative process. The initial set of guidelines were distilled from Wang et al.'s \cite{cheryl_study} recent work about challenges in addressing OSS usability. Then the guidelines were revised as more information was gleaned through the user studies of preliminary prototypes (see Section \ref{section:user_study}). When creating these guidelines, we particularly concentrated on the aspects that can facilitate asynchronous participatory design in the OSS context. The final design guidelines for the system are summarized in Tables \ref{tab:designer_goals} and \ref{tab:enduser_goals}, as well as the corresponding challenges identified either by Wang et al~\cite{cheryl_study} or our user study.

\subsection{Prototype Design Process}
\label{section:designprocess}

\begin{figure*}[t]
    \centering
    \begin{subfigure}[b]{0.32\textwidth}
        \frame{\includegraphics[width=\textwidth]{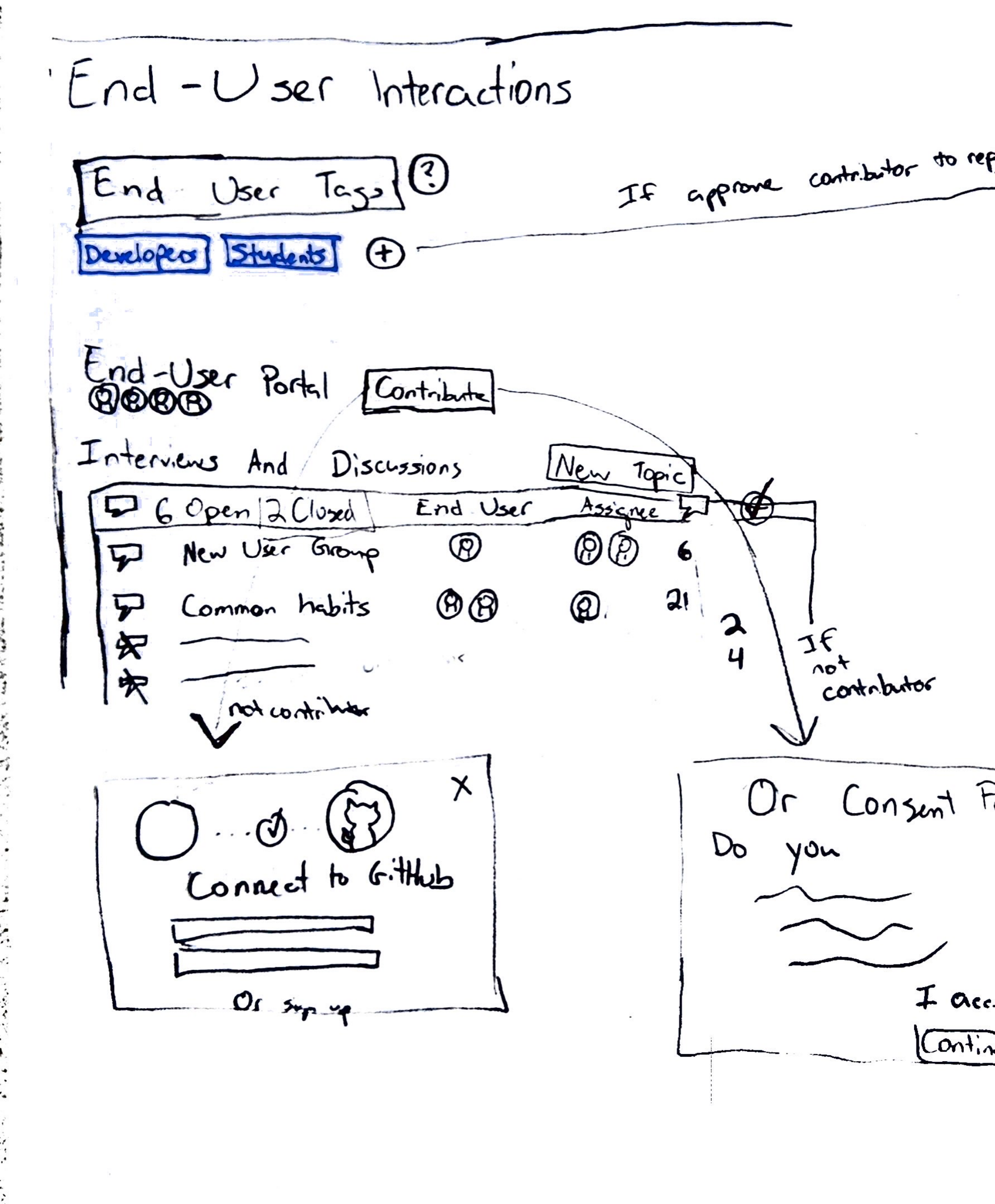}}
        \caption{End user interactions with user tags\\ and user collaboration portal}
        \label{fig:userinvolvement}
    \end{subfigure}\hspace{0.01\textwidth}
    \begin{subfigure}[b]{0.32\textwidth}
        \frame{\includegraphics[width=\textwidth]{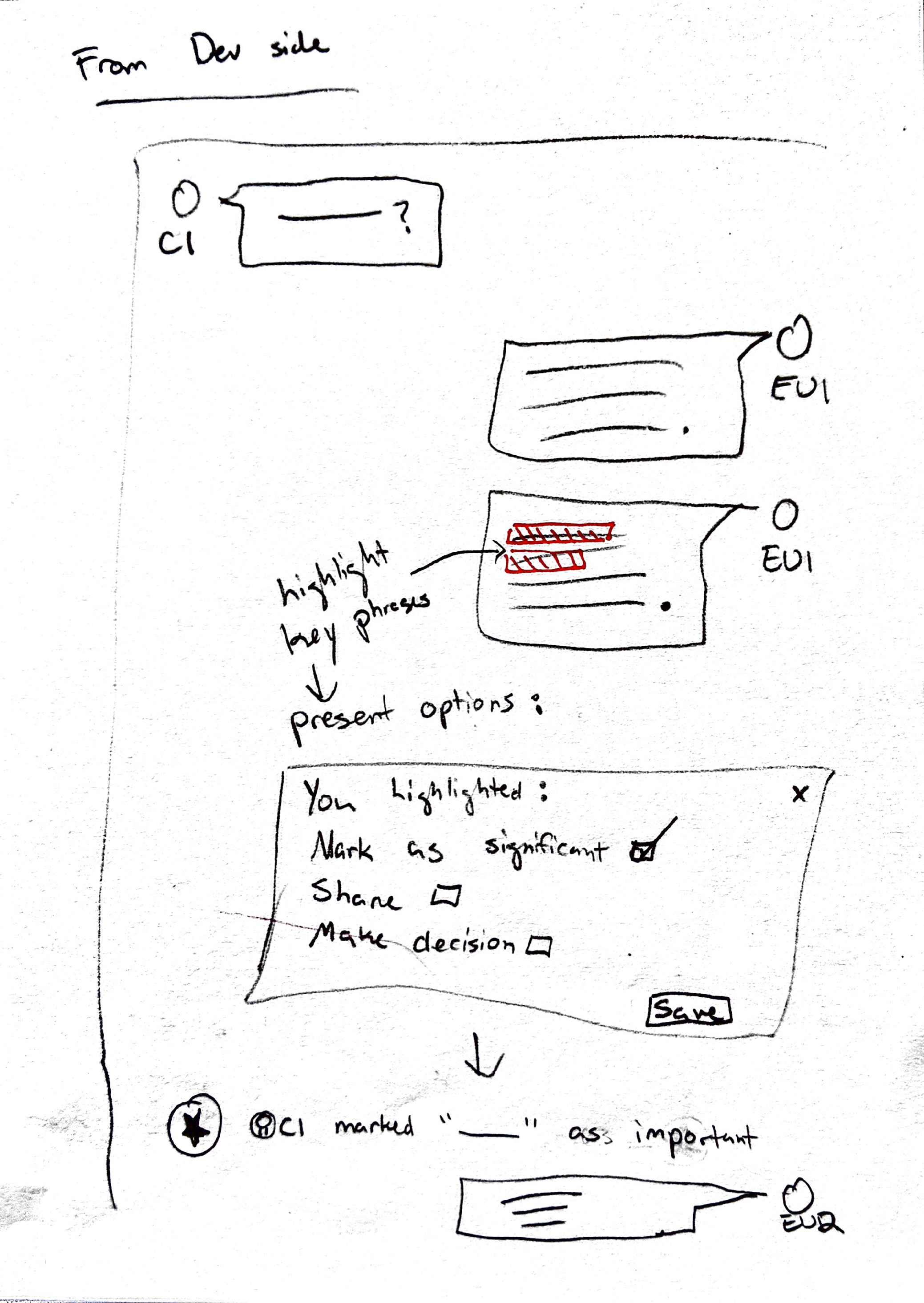}}
        \caption{Communication messages design\\ with pipeline integration}
        \label{fig:pipeline}
    \end{subfigure}\hspace{0.01\textwidth}
    \begin{subfigure}[b]{0.32\textwidth}
        \frame{\includegraphics[width=\textwidth]{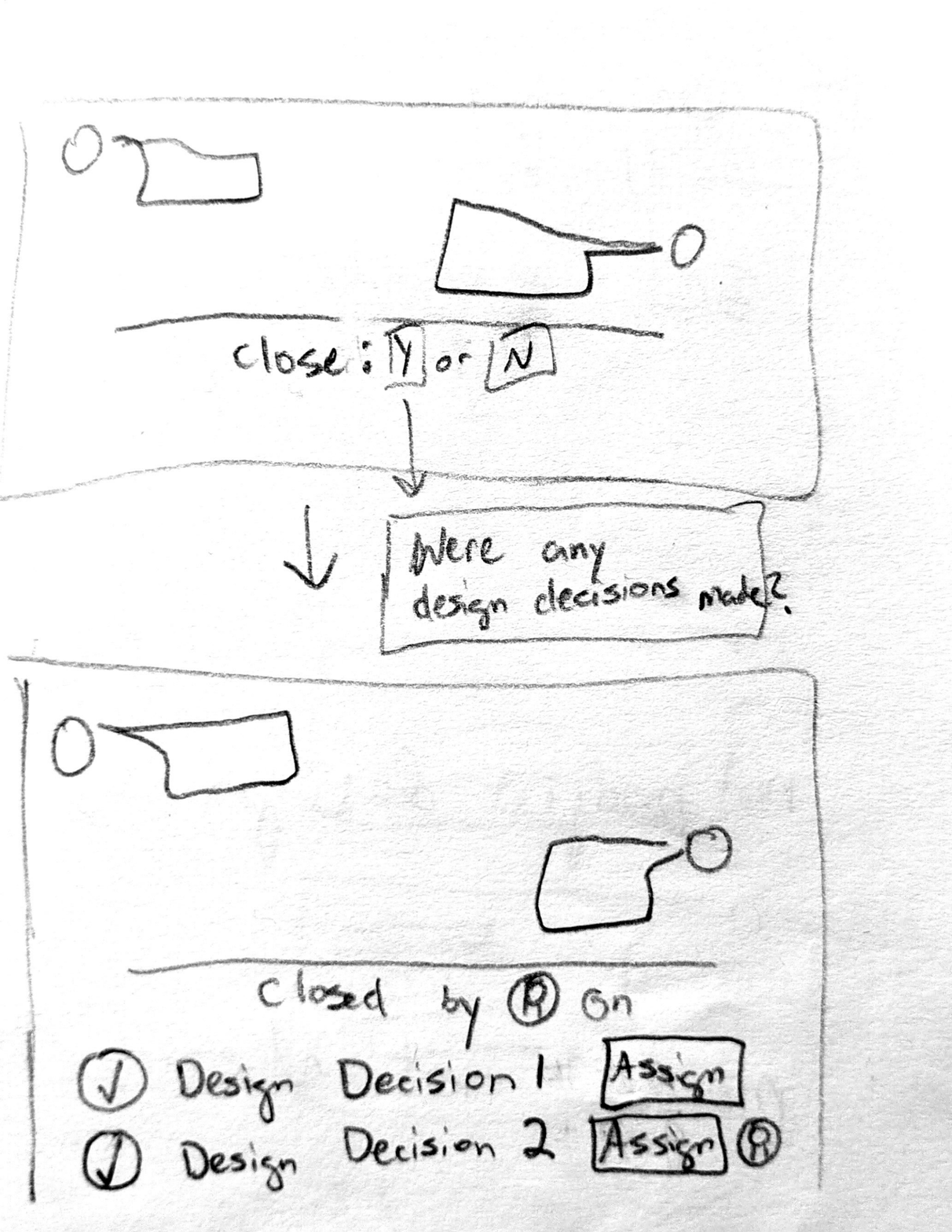}}
        \caption{Design decision artifact creation and assignment to contributors}
        \label{fig:designartifact}
    \end{subfigure}
    \caption{Initial Sketches of the System Design}
    \label{fig:sketches}
    \Description[Three hand-drawn UI sketches]{Hand-drawn sketches of the user interface design for three components of the system}
\end{figure*}

We design the preliminary system that can be integrated into existing OSS hosting services, such as GitHub\footnote{https://github.com}. 
While those services include various asynchronous collaboration features that are widely adopted by OSS communities, they generally fall short of delivering the same level of benefits for asynchronous, social \textit{design}.

Designing the prototype of the system had three overlapping phases. First the research team brainstormed, sketched, and discussed the system design based on the initial design guidelines. Some initial sketches are shown in Figure~\ref{fig:sketches}. These sketches focused on the following main features that reflected the design guidelines.
\begin{itemize}[leftmargin=12pt]
    \item End User Involvement (see Figure~\ref{fig:userinvolvement}) - This feature captures the guidelines D1, D5, EU1, EU2, EU4, and EU5 and provides a portal, separated from the rest of the code management system, for designers to define target end users and their roles, for end users to sign-up for collaborations, and facilitate direct and intuitive interactions between designers and end users.
    \item Pipeline Integration (see Figure~\ref{fig:pipeline}) - This feature captures the guidelines D2, D4, and EU3 and provides ability to integrate the proposed system within existing OSS development pipelines through the creation of associated `Issues' from discussions of usability concerns with end users.
    \item Collaborative Design Artifact Generation (see Figure~\ref{fig:designartifact}) - This feature captures the guidelines D3 and EU4 and facilitates the co-creation and feedback collection of design artifacts with the end users. This feature can be utilized to achieve \textit{Pipeline Integration} and provides transparency to both OSS end users and other members of the project team (e.g. developers and maintainers).
\end{itemize}
Second, we evolved the sketches with initial feedback from OSS designers and end users. During these initial user studies, two personas of the target users were created to guide further design. Due to the novelty of the system, we chose to develop detailed personas in parallel to the prototype exploration so that we could perform iterations as needed as new information was gleaned about the needs of the target user groups while actually examining the proposed system design. In the third phase of prototype design, we built an interactive prototype of the tool using Framer (www.framer.com) based on the evolved sketches and guided by the personas. This prototype is to be used for further user feedback through another round of studies. In the following sections, we detail the user study we conducted in the second phase of design, present the personas we created, and then present the latest version of the prototype.

\section{User Study}
\label{section:user_study}
We conducted a preliminary user study with two OSS designers and two OSS end users to get feedback on the initial sketches and to iterate the early versions of the prototype. This user study was approved by the Ethics Board of the institution of the researchers.

\subsection{Methods}
Participants were were recruited from both personal networks and the Open Source Design forum; Open Source Design is a community of designers and developers committed to improving design in OSS\footnote{https://discourse.opensourcedesign.net/}. The characteristics of the participants are summarized in Table~\ref{tab:participants}. Among the four participants, one was non-binary, one was female, and two were male. Both OSS designers had considerable experience with OSS design projects; all their latest design projects were utilized by non-technical end users.

\begin{table*}[t]
    \centering
    \caption{User Study Participants' Characteristics}
    \resizebox{\textwidth}{!}{%
    \begin{tabular}{cccc} \hline
        \textbf{Participant ID} & \textbf{Target Group}  & \textbf{Experience} & \textbf{Current Occupation}\\ \hline
         P1D & OSS Designer & Decades working in OSS in the public health sector & Design Director/Lecturer \\
         P2D & OSS Designer & About one decade working in OSS as lead designer in social enterprises & Design Lead \\
         P3EU & OSS End User & Casual OSS end user of Mozilla Firefox, Audacity & Project Coordinator Assistant\\
         P4EU & OSS End User & OSS end user Notepad++, Audacity, Gimp & PhD Candidate/Lecturer \\
         \hline
    \end{tabular}%
    }
    \label{tab:participants}
\end{table*}

We used two different sets of interview questions for the designers and end users, respectively, during the one-on-one video conferencing calls. The OSS Designer interview focused on the participants' (a) experience and roles, (b) current tools and processes used to achieve OSS participatory design, (c) current challenges faced in making design decisions, and (d) open ended feedback on the sketches. Similarly, the OSS End User interview included specific questions regarding (a) general experience and roles, (b) utilizing OSS, (c) becoming involved in OSS development and challenges that were faced, and (d) open ended feedback on the sketches.

The interviews were recorded and then inductively coded to analyze themes~\cite{guest2011applied}. Following the analysis, a summary of the main insights was drafted and used to (1) modify the design guidelines of the proposed system, (2) evolve the design of the proposed system, and (3) create the personas to capture the user needs.

\begin{figure*}[t]
    \begin{subfigure}[b]{0.48\textwidth}
        \includegraphics[width=\textwidth]{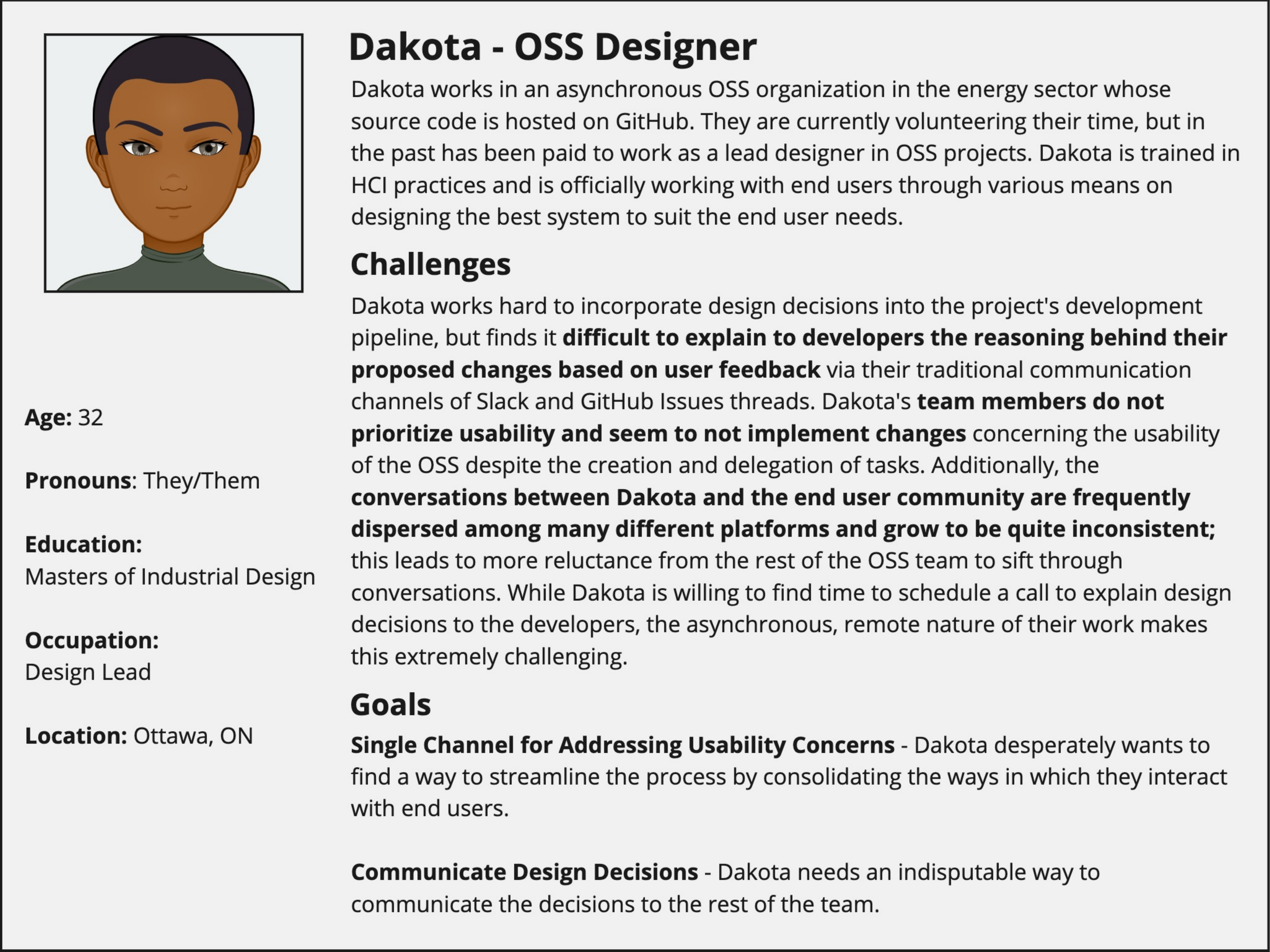}
        \caption{OSS Designer Persona Dakota}
    \end{subfigure}\hspace{0.01\textwidth}
    \begin{subfigure}[b]{0.48\textwidth}
        \includegraphics[width=\textwidth]{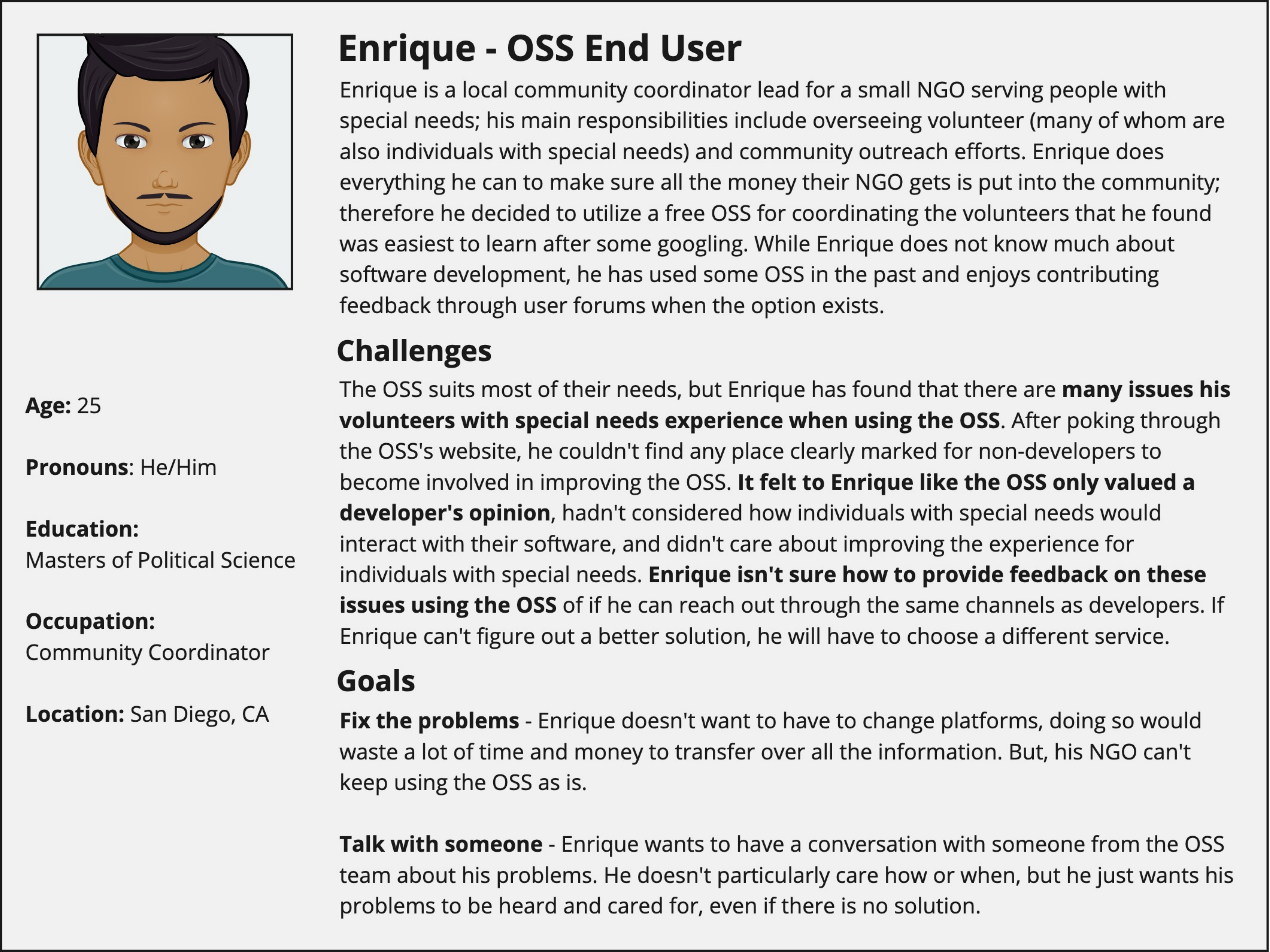}
        \caption{OSS End User Persona Enrique}
    \end{subfigure}
    \caption{Personas Created Through the Study}
    \label{fig:personas}
    \Description[One designer persona and one end user persona]{One designer persona and one end user persona, each includes the demographics, a brief introduction, a description of challenges, and a list of goals.}
\end{figure*}

\subsection{Results}
In this section, we summarize the key insights from our user studies. We begin with the insights from the interviews with the OSS designers:

\begin{enumerate}[label={\textit{Designer Insight \arabic*:}},labelindent=-30pt,itemindent=46pt]
    \item There are no current standards for addressing usability concerns with OSS end users and community members. P2D mentioned that each project they worked on utilized different design methods to practice participatory design. Both P1D and P2D utilized many different platforms and tools to practice participatory design (e.g., Figma, Airtable, screen recorded demos, Twitter, emails, Slack, Discourse, video conferencing, etc.).

    \item Two major consequences resulting from \textit{Designer Insight 1} were bottlenecks in the pipeline from addressing usability and difficulties in communicating ideas to different groups of people on various tasks. Especially with the asynchronous culture in which P2D works, there wasn't a set process for providing design artifacts and necessary changes on the software design were not always accomplished.

    \item The OSS designers positively received the idea of moving usability discussions and interactions between designers and end users out of `Issues', which is already an overloaded feature on OSS platforms~\cite{argulens}. P2D mentioned that the developers they collaborated with frequently ignored issues flagged as usability concerns and argued with P2D against proposed changes resulting from end user discussions. P2D believed that decluttering the `Issues' would lead to less aggravation of developers, potentially resulting in a mindset shift towards more receptive to usability issues.

    \item Participants discussed that integrating and tracing the visual design artifacts in the OSS collaboration platforms such as GitHub would be helpful. P1D said: ``\textit{Feels like a no-brainer to have a more visual asset visualizer... on GitHub. Our team would greatly benefit in SEEING artifacts that are inherently visual.}''

    \item Participants also emphasized the importance to ``\textit{keep the core GitHub repo concepts}'' where ``\textit{the Design Layer should not "interfere" with core engineering activities}'' (P1D). They expressed that any additional feature should be consistent with the existing engineering culture and use of the collaboration platforms. 
\end{enumerate}

Interviews with the OSS End Users brought to light many important distinctions for our system design that had previously gone unconsidered.
Listed below are several key insights.

\begin{enumerate}[label={\textit{Eng User Insight \arabic*:}},labelindent=-30pt,itemindent=46pt]

\item From the end user's perspective, a major barrier to contributing to an OSS was not knowing how. While P4EU had found the GitHub repository for one OSS through their website, they expressed difficulty in understanding how GitHub works and how to word their comments to ensure mutual understanding. For this, P4EU stated GitHub was overwhelming and intimidating.
P3EU felt an ideal way to be included in the design would be through surveys, feedback, or being contacted for participating in a conversation.
\item To address the above issue, the onboarding pipe-line for end users to the proposed system needs to prioritize simplicity. After viewing the initial sketches, P3EU felt overwhelmed by the onboarding pipeline starting from landing on the main page of a GitHub repository while P4EU expressed minimizing the initial steps an OSS end user would need to perform to provide feedback is crucial to maximize end user involvement (such as not requiring an account, see Figure \ref{fig:userinvolvement}).

\item It is important for the tool design to indicate to OSS end users that they are welcome and that their opinions/experiences are valid and helpful. Several key features in the sketches can help to accomplish this, as indicated by P4EU, including the `end user tag' features (see Figure \ref{fig:userinvolvement}) and the informality of the conversation format (see Figure ~\ref{fig:pipeline}). Following on this, P4EU also discussed the importance in terminology used throughout the system design; in many cases, P4EU felt that the tone of a selected word would deter them from participating in feedback. For example, the use of `Contribute' (see Figure \ref{fig:userinvolvement}) was received negatively.

\item Both end user participants felt that it is valuable to see that concrete actions are taken by developers based on their conversations on usability concerns. They valued the feature of linking design artifacts to the discussions (see \ref{fig:designartifact}), considering that it would enhance their experience in collaborating and would motivate them for continued efforts.

\item The two interviews illuminated necessary updates to the original target audience description for OSS End Users. Particularly, this group needs to possess characteristics demonstrative of early adopters and innovators of technology \cite{early_adopters, oss_user_definition}. This emphasis is important because these user groups traditionally are the one's which provide critical usability feedback in non-OSS contexts.

\end{enumerate}

\section{Personas}
\label{section:personas}
During the exploration of the initial design of the system, we developed two personas, one for OSS designers and one for OSS end users, and iterated them based on the user study results (see Figure~\ref{fig:personas}). The OSS Designer persona is Dakota, a lead designer who struggles to manage the various platforms for interacting with their end user community and communicating the decision outcomes from those interactions to the rest of their OSS team. The OSS End User persona is Enrique, a community coordinator for a local non-governmental organization (NGO) who is using an OSS to coordinate volunteers; Enrique has experienced issues with the OSS when coordinating volunteers with special needs and he faces difficulties communicating these issues with the OSS team. We envision that these personas can support the design of this and similar systems that aim at facilitating OSS user-designer communication. We will use them in the next iterations of system design.

\begin{figure*}[t]
    \centering
    \includegraphics[width=\textwidth]{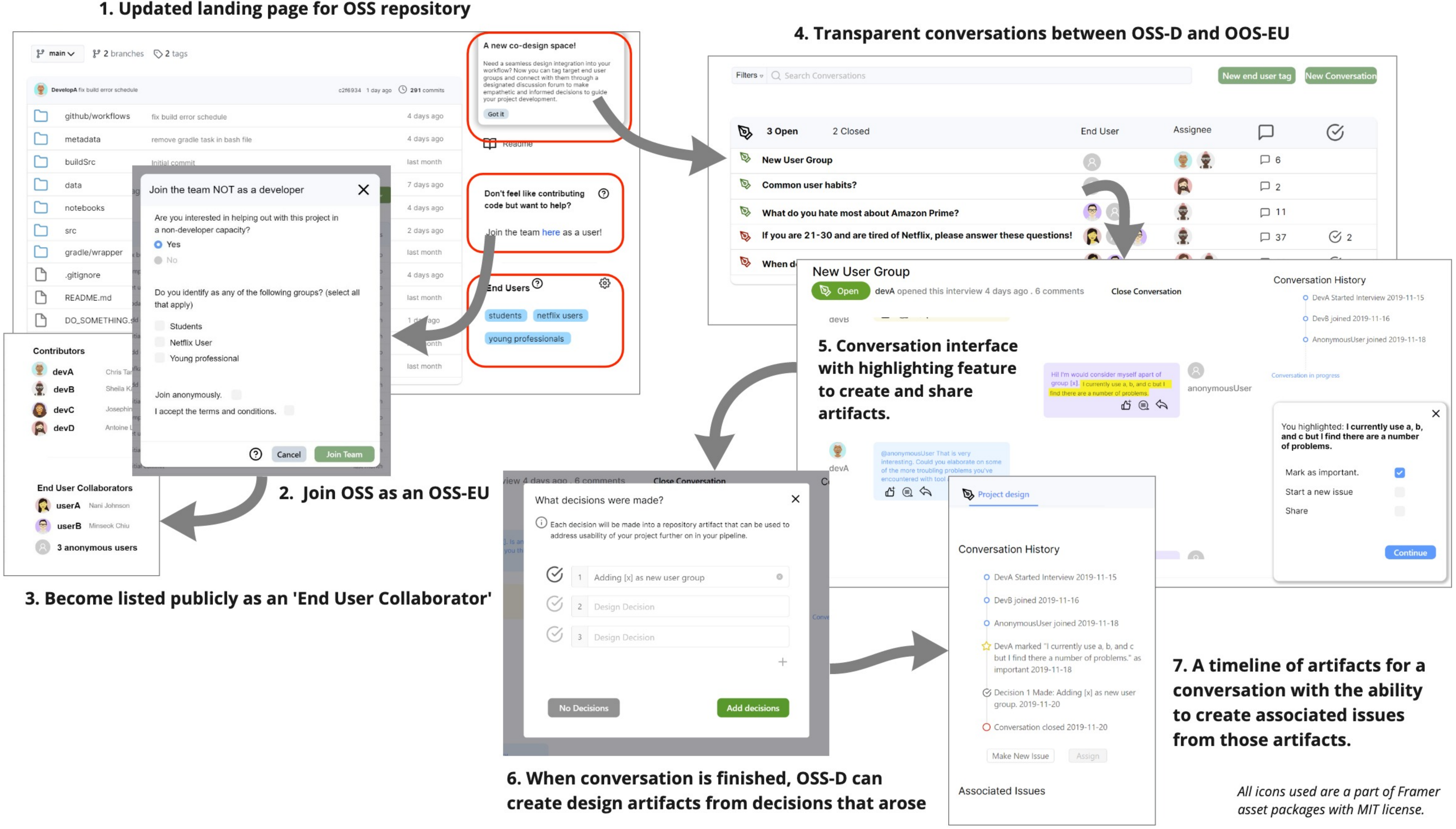}
    \caption{Overview of the Prototype System Design}
    \label{fig:prototype_screenshot}
    \Description[A set of UI mockups organized in a workflow that includes seven steps]{UI mockups organized in a workflow that includes seven steps: 1. landing page for OSS repository; 2. end users join the OSS project by selecting their user group; 3. end users become listed publicly as collaborators; 4. a UI for managing the transparent conversations between OSS designers and the end users; 5. conversation UI with a highlighting feature to create and share artifacts; 6. when conversation is finished, designers can create design artifacts from decisions that arose; and 7. a timeline of artifacts for a conversation with the ability to create associated issues.}
\end{figure*}

\section{Latest Version of System Design}
The user feedback on the sketches resulted in a few notable changes to the prototype design. The workflow of the latest version of the interactive prototype is summarized in Figure~\ref{fig:prototype_screenshot}.
To emulate a more familiar conversational interaction for the OSS end users following insights from P4EU, the design of interaction between OSS designers and end users resembles platforms such as Slack and other asynchronous messaging platforms with features to react to a comment and start a thread (Guidelines D1, D5, EU1, EU2, EU4, EU5); see Figure~\ref{fig:prototype_screenshot} step 5.
The addition of an `End User Collaborator' section underneath the standard `Contributors' list for a repository should encourage activity due to visible recognition of their efforts and display of OSS end users as valued members of the OSS community (Guidelines D5, EU4, EU5); see Figure~\ref{fig:prototype_screenshot} step 3.
Clear buttons to create new software issues from resulting design artifacts of a conversation now incorporated for seamless integration into the existing pipeline of an OSS project; this feature provides the ability to directly link the relevant artifacts in an issue and indicate in the conversation who has been assigned to work on the new issue to ensure transparency and traceability (Guidelines D2, D3, D4, D5, EU4, EU5).
The system design is also guided by several usability heuristics (e.g. consistency, minimalist design, and flexibility~\cite{nielsen_2020}) to achieve an effective and efficient interaction.

\section{Discussion and Conclusion}
Overall, the feedback received during the user studies were overwhelmingly supportive towards the goals of understanding key barriers experienced by OSS end users taking part in participatory design efforts and of creating a tool to facilitate successful, asynchronous participatory design.
Our user study highlights the need for a standardized method to support participatory design interactions between OSS end users and designers.
While the current prototype still needs to undergo thorough usability testing to demonstrate if it successfully achieves these goals, the user tests that have been conducted thus far support the creation of a singular tool that follow the guidelines and the three core features established in Sections ~\ref{section:designguidelines} and \ref{section:designprocess}.

Despite overall positive feedback on the sketches and early prototype, the proposed design poses some challenges illuminated by the study participants. Participant P2D indicated that every time a new platform for communication is introduced to their end user community, switching costs are incurred as not all end users ``\textit{want}'' to learn a new technology. (In this example, P2D explained the challenges with moving from Discourse to Slack where now both are active.) It is important to further explore this dimension with the the next round of user studies. In particular, exploring how to effectively integrate this system with existing tools and workflows would be important. Furthermore, the current design of the system for a new end user to join or start a conversation solely through the OSS repository might prove to go against some of the usability concerns, as was indicated by both OSS end user participants' ambivalence towards the GitHub-influenced design. A possible change to the system might need to include a dedicated OSS end user-facing portal or interface to make it as easy as possible for diverse user bases to join participatory design efforts.

In summary, this paper contributes a set of design guidelines and two personas through an user-centered exploration of a prototype tool for OSS designers and end users in collaboratively addressing usability concerns. Our preliminary user study confirmed a need for a standardized method to facilitate user-designer interactions that prioritizes efficient communication of usability concerns to other OSS team members. Needs also emerged for an interface design that uses concepts and terminology that is inclusive of OSS end users with diverse technical backgrounds. Our design guidelines and personas focused on capturing these and other previously identified needs for promoting OSS usability. We are currently iterating the prototype with additional user studies to improve the design.

\section{Acknowledgement}
We thank our participants for their time and valuable feedback. This work was partially supported by the Discovery Grant of the Natural Sciences and Engineering Research Council of Canada.

\bibliographystyle{ACM-Reference-Format}
\bibliography{bibliography}

\end{document}